\documentclass[11pt,preprint]{aastex}

\def\t0{\theta_{\circ}}

\def\be{\begin{equation}}
\def\en{\end{equation}}

\def\msun{M_{\sun}}
\def\rsun{R_{\sun}}
\def\lsun{L_{\sun}}
\def\msunyr{M_{\sun} \, yr^{-1}}
\def\kms{\rm \, km \, s^{-1}}
\def\mdot{\dot{M}}

\def\h2{H$_2$}

\begin{document}

\title
{NUV Excess in Slowly Accreting T Tauri Stars: Limits Imposed by Chromospheric Emission*\footnotetext[1]{This paper includes data gathered with the 6.5 meter Magellan Telescopes located at Las Campanas Observatory, Chile.}}

\author{Laura Ingleby\altaffilmark{1}, Nuria Calvet\altaffilmark{1}, Edwin Bergin\altaffilmark{1}, Gregory Herczeg\altaffilmark{2}, Alexander Brown\altaffilmark{3}, Richard Alexander\altaffilmark{4}, Suzan Edwards\altaffilmark{5}, Catherine Espaillat\altaffilmark{6}, Kevin France\altaffilmark{3}, Scott G. Gregory\altaffilmark{7}, Lynne Hillenbrand\altaffilmark{7}, Evelyne Roueff\altaffilmark{8}, Jeff Valenti\altaffilmark{9}, Frederick Walter\altaffilmark{10}, Christopher Johns-Krull\altaffilmark{11}, Joanna Brown\altaffilmark{12}, Jeffrey Linsky\altaffilmark{13}, Melissa McClure\altaffilmark{1}, David Ardila\altaffilmark{14}, Herv{\'e} Abgrall\altaffilmark{8}, Thomas Bethell\altaffilmark{1} , Gaitee Hussain\altaffilmark{15},  Hao Yang\altaffilmark{12}
}

\altaffiltext{1}{Department of Astronomy, University of Michigan, 830 Dennison Building, 500 Church Street, Ann Arbor, MI 48109, USA; lingleby@umich.edu, ncalvet@umich.edu}
\altaffiltext{2}{Max-Planck-Institut f\"ur extraterrestriche Physik, Postfach 1312, 85741 Garching, Germany; gregoryh@mpe.mpg.de}
\altaffiltext{3}{Center for Astrophysics and Space Astronomy, University of Colorado, Boulder, CO 80309-0389, USA}
\altaffiltext{4}{Department of Physics \& Astronomy, University of Leicester, University Road, Leicester, LE1 7RH, UK}
\altaffiltext{5}{Department of Astronomy, Smith College, Northampton, MA 01063, USA}
\altaffiltext{6}{NSF Astronomy and Astrophysics Postdoctoral Fellow.  Harvard-Smithsonian Center for Astrophysics, 60 Garden Street, MS-78, Cambridge, MA 02138, USA}
\altaffiltext{7}{California Institute of Technology, Department of Astrophysics, MC 249-17, Pasadena, CA 91125, USA}
\altaffiltext{8}{LUTH and UMR 8102 du CNRS, Observatoire de Paris, Section de Meudon, Place J. Janssen, 92195 Meudon, France}
\altaffiltext{9}{Space Telescope Science Institute, 3700 San Martin Dr, Baltimore MD 21218, USA}
\altaffiltext{10}{Stony Brook University, Stony Brook NY 11794-3800, USA}
\altaffiltext{11}{Department of Physics \& Astronomy, Rice University, Houston, TX 77005, USA}
\altaffiltext{12}{Harvard-Smithsonian Center for Astrophysics, 60 Garden Street, MS-78, Cambridge, MA 02138, USA}
\altaffiltext{13}{JILA, University of Colorado and NIST, 440 UCB Boulder, CO 80309-0440, USA}
\altaffiltext{14}{NASA Herschel Science Center, California Institute of Technology, Mail Code 100-22, Pasadena, CA 91125, USA}
\altaffiltext{15}{ESO, Karl-Schwarzschild-Strasse 2, 85748 Garching bei M\"{u}nchen, Germany}

\begin{abstract}
Young stars surrounded by disks with very low mass accretion rates are
likely in the final stages of inner disk evolution and therefore
particularly interesting to study.  We present ultraviolet (UV)
observations of the $\sim$5--9 Myr old stars RECX-1 and RECX-11, obtained with the Cosmic Origins
Spectrograph (COS) and Space Telescope Imaging Spectrograph (STIS) on the Hubble Space Telescope (HST), as well as optical and near infrared
spectroscopic observations.  The two stars have similar levels
of near UV emission, although spectroscopic evidence indicates
that RECX-11 is accreting and RECX-1 is not. The line profiles of
H$\alpha$ and He I $\lambda$10830 in RECX-11 show both broad and narrow
redshifted absorption components that vary with time, revealing the
complexity of the accretion flows.  We show
that accretion indicators commonly used to measure mass accretion
rates, e.g. $U$ band excess luminosity or the Ca II triplet line
luminosity are unreliable for low accretors, at least in the middle
K spectral range. Using RECX-1 as a template for the intrinsic level
of photospheric and chromospheric emission, we determine an upper
limit of $3\times10^{-10}\;\msunyr$ for RECX-11. At this low accretion
rate, recent photoevaporation models predict that an inner hole should
have developed in the disk. However, the spectral energy distribution
of RECX-11 shows fluxes comparable to the median of Taurus in the
near infrared, indicating that substantial dust remains.  Fluorescent H$_2$ emission lines formed in the innermost disk are observed
in RECX-11, showing that gas is present in the inner disk, along with the dust.

\end{abstract}

\keywords{Accretion, accretion disks, Stars: Circumstellar Matter, Stars: Pre Main Sequence}

\section{ Introduction}
\label{intro}
It is now accepted that 
classical
T Tauri stars (CTTS{\footnote {In this paper we use the terms
    classical and weak-line T Tauri stars (CTTS and WTTS) to indicate
    accreting and non-accreting T Tauri stars, respectively, following
    \citet{white03}.})
are accreting mass from their
disks via magnetospheric accretion.  
In this process,
the stellar magnetic field truncates the circumstellar
  disk and channels gas near the truncation radius onto the stellar
  surface, where an accretion shock forms.  Gas heated by the shock emits
  energetic photons which, after re-processing in the accretion
  streams, irradiate the inner disk with ultraviolet (UV) and X-ray
  photons \citep{calvet98,ardila09,gunther07}.  This energetic
  emission plays a key role in driving disk chemistry and may be the
  most efficient mechanism for dispersing the disk gas
  \citep{najita10,pinte10,bethell09,gorti09}.  The rate at which
  material is accreted is an important input in models of
  circumstellar disks, affecting the surface density and
  inner disk heating \citep{dalessio06}, so accurate accretion rates
  are essential for describing the evolution of both gas and dust.
  Accretion rates decrease with age \citep{calvet05}, so objects
with very low
  accretion rates are particularly interesting for studying the final
  stages in the evolution of disks.

The spectrum produced by the accretion shock on the
stellar surface peaks in the UV, so UV
flux
excesses over the stellar fluxes are the most direct measure of the accretion luminosity
($L_{acc}$) from which the mass accretion rate, $\mdot$, can be
derived (\S\ref{shock}).  
However, CTTS 
have active chromospheres which produce UV emission in excess of
the main sequence standards \citep{houdebine96,franchini98}.  
The chromospheric contribution to the measured
excess is expected to be small when compared to the excess
due to accretion in CTTS with typical values
of the mass accretion rate, $\mdot \sim 10^{-8} \, \msunyr$
\citep{gullbring98}.
However,
as the disk evolves, the mass accretion rate
and accretion luminosity decrease, and so do the
accretion-powered UV fluxes.
In contrast, 
indicators of chromospheric activity, e.g. the X-ray luminosity, stay approximately
constant in the age range 1 - 10 Myr during which disks
evolve significantly \citep{ingleby11}. Therefore,
the chromospheric contribution is expected
to become more important in determining the total UV flux
as the disk evolves and at some point will become
the dominant contributor. From this point on, the
UV excess is no longer measuring accretion.

In most determinations of accretion luminosity in CTTS to date,
the excess fluxes in the UV have been measured
using main sequence standards
as templates
for the intrinsic stellar emission.
A better template for
the stellar emission of an accreting T Tauri star is a non-accreting T
Tauri star (or WTTS), which is expected to have a comparable level of
chromospheric activity as a CTTS.  
However, UV observations of WTTS with enough
sensitivity do not exist \citep{valenti03},
so it has not been possible to determine
typical levels of chromospheric emission,
which in turn are important to set up 
realistic limits to the accretion luminosities
that can be determined from UV observations.
In this paper we
address these issues by comparing
the UV fluxes of a WTTS, RECX-1,
with those of a very low accretor, RECX-11, 
and find that the NUV fluxes are dominated by chromospheric emission;
this allows us to 
place
limits on the lowest accretion luminosites detectable in the UV.

Even for chromospheric-dominated UV fluxes,
alternative indicators may be used to
identify whether a young star is accreting. 
A commonly used accretion indicator is
the H$\alpha$ line profile.  H$\alpha$ is observed in emission in all
T Tauri stars but a large velocity width or line asymmetry indicates
accretion, whereas a narrow symmetric line profile is characteristic
of a non-accreting star \citep{muzerolle01,lima10,white03}.  
In addition, redshifted absorption components superimposed on the emission
line profiles are direct indication of infalling material and 
can be explained in terms of magnetospheric accretion \citep{muzerolle98,muzerolle01}.
The He I
line at 10830 {\AA} can also be used to diagnose accretion properties
in T Tauri stars.
In CTTS this line shows deep blueshifted and redshifted absorption components
in its profile, which have been interpreted as 
formed in outflows and accretion streams
\citep{edwards06,fischer08}.
Other indicators of accretion used in the literature are the forbidden
lines of [OI] 6300 and 6363 {\AA}, which are present only in accreting stars
\citep{hartigan95}, and the Ca II triplet lines, whose luminosites 
correlate with accretion luminosity \citep{calvet04}.
In this paper we analyze these accretion indicators in high
resolution spectra of RECX-11 and RECX-1.
We find conclusive evidence for accretion in RECX-11 and
not in RECX-1 in our
high resolution observations of H$\alpha$ and He I $\lambda$10830,
but not in other lines.
Our analysis allows us to assess
which accretion indicators are appropriate for detecting accretion
at very low levels.

The indicators discussed above are directly tracing material in the
accretion shock or flows, but another indirect method of determining
whether a source is accreting or not involves gas in the inner
circumstellar disk.  While in the past circumstellar gas was difficult
to observe, new observations have made gas detections possible
\citep{bary08,bitner08,carr08,pascucci09a}.  Of particular interest are
observations of hot gas very close to the star, near the
magnetospheric truncation radius.  FUV observations have shown that CO
and \h2 at temperatures near 2500 K are present in the disk
\citep{france11,france10,herczeg04,herczeg02}.  
\citet{ingleby09}
analyzed a sample of FUV spectra of T Tauri stars and found that \h2
was present in the inner 1 AU of all accreting sources but not 
in non-accreting sources. 
FUV gas detections therefore indicate
more than the presence of gas, but actually show whether a source is
still accreting or not.
These results were based on ACS low resolution observations;
the COS observations presented in this paper
allow us to confirm this finding using
spectrally resolved \h2 lines.

This paper is organized as follows. In \S\ref{obs} we present the UV,
optical and IR data discussed in this paper.  In \S\ref{evidence} we
discuss which commonly used accretion indicators are valid at low
levels of accretion
and which are unreliable.  In \S\ref{smallmdot} we 
estimate limits for the minimum $\mdot$
that can be determined from the UV.
Finally, in \S\ref{discussion} we discuss the
limitations of current accretion rate determinations in view of our
results, as well as the implications for disk evolution models.

\section{Observations}
\label{obs}

\subsection{The Targets}
Our targets are the young stars RECX-1 and RECX-11 in the $\eta$
Chamaeleontis group, the nearest open cluster at only 97 pc with an
age of 5-9 Myr \citep{mamajek99,mamajek00,luhman04}.  With $A_V\sim0$,
this region has little extinction \citep{luhman04}, making it ideal
for UV observations as corrections for reddening may introduce large
uncertainties.  The sources have similar spectral types, and were
classified as K4 by \citet{mamajek99} and K5-K6 by \citet{luhman04}.
Sources in this region have also been well characterized in the
infrared with broad band $J-L$ photometry \citep{lyo04},
\emph{Spitzer} IRAC and MIPS photometry and Infrared Spectrograph
(IRS) spectroscopy \citep{sicilia09}.  The properties of each source
are summarized in Table \ref{tabprop}.

\begin{figure}
\plotone{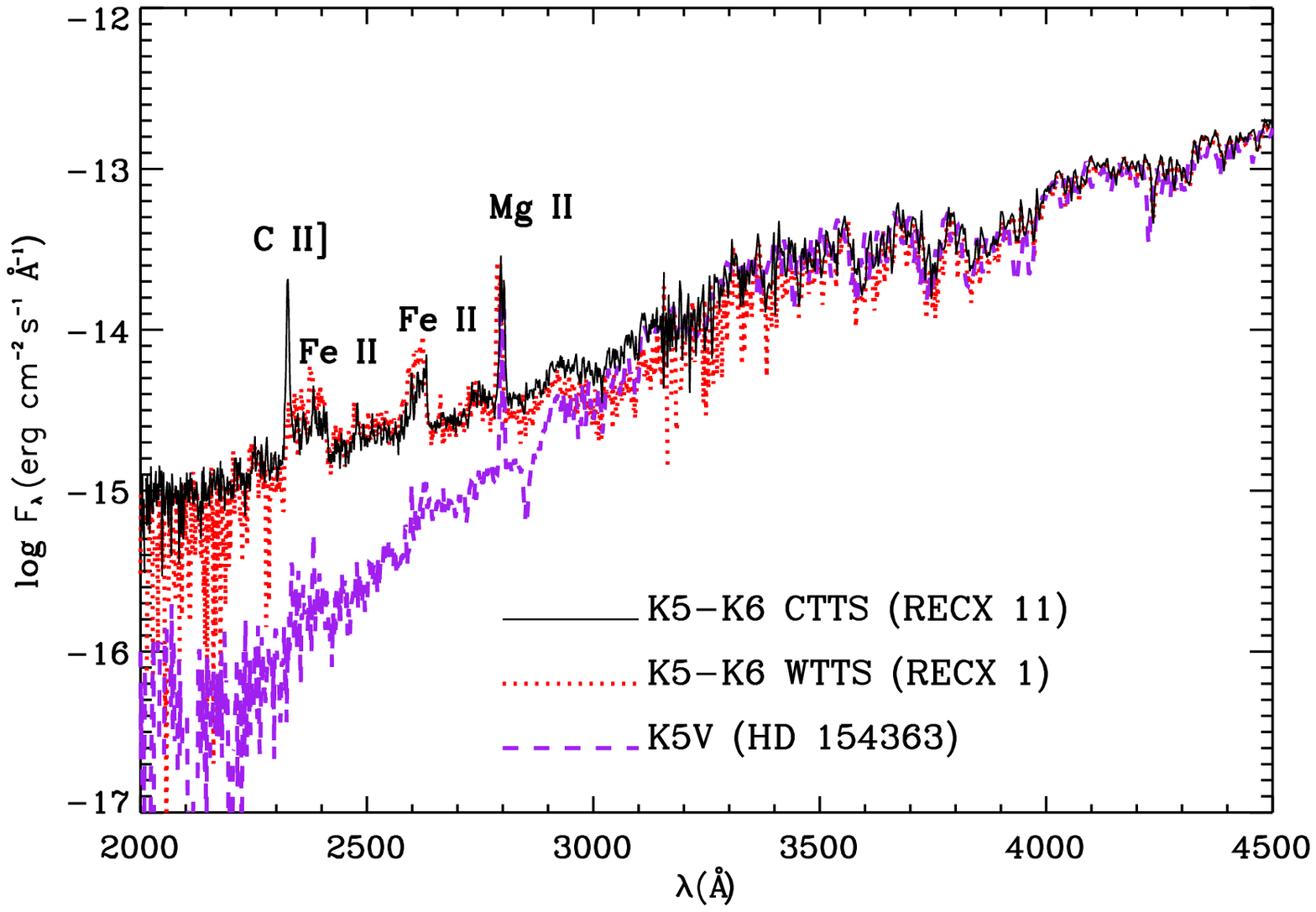}
\caption{STIS spectra of the WTTS RECX-1 (both north and south components,
see \S\ref{hst}), the CTTS RECX-11 and the dwarf standard star, HD 154363.
All have K5--K6 spectral types but RECX-1 is no longer accreting while
RECX-11 has ongoing accretion \citep{lawson04, sicilia09,jay06}.  The spectra of the RECX-1
and RECX-11 are remarkably similar, except for the stronger C II]
$\lambda$2325 line in the CTTS.  The Mg II $\lambda$2800 doublet strength
is comparable in the three sources ($\S$\ref{smallmdot}).}
\label{stis}
\end{figure}

\subsection{HST Observations}
\label{hst}

Observations of RECX-1 and RECX-11 were obtained between 2009 December
and 2010 January with COS and STIS on HST in GO Program 11616 (PI:
Herczeg).  
STIS NUV observations used the MAMA detector and the G230L
grating providing spectral coverage from 1570 to 3180 {\AA} with
R$\sim$500-1000.  Optical observations were completed during the same
orbit as the NUV using the G430L grating which covers 2900-5700 {\AA}
with R$\sim$530--1040, resulting in almost simultaneous NUV to optical
coverage with STIS.  

The low-resolution STIS spectra were calibrated with custom written
IDL routines following the procedures described in the STIS data
handbook.  The wavelengths were calibrated from the location of
identified emission lines within the spectrum, and fluxes were calibrated
from spectra of WD 1337+705 in the NUV and HIP 45880 in the optical.
The flux calibration also includes a wavelength-dependent aperture
correction.

COS observations of RECX-11 were presented in \citet{france11}, and we
discuss them here together with observations for RECX-1.  The COS FUV
observations were taken with the G160M and G130M gratings and were
completed within 3--4 hours of the STIS observations.  The combination
of the two gratings provides FUV coverage from 1150 -- 1775 {\AA} with
R$\sim$16,000 -- 18,000.  The log of the observations is given in
Table \ref{tabobs}.

The COS spectra were processed by the standard CALCOS calibration
pipeline.  The individual spectral segments, after obtaining different
wavelength settings to minimize instrumental fixed pattern effects and
to provide full wavelength sampling, were combined using the IDL
coaddition procedure described by \citet{danforth10}. This procedure
coaligns the individual exposures and performs an exposure-weighted
interpolation onto a common wavelength grid.

In our COS NUV acquisition image, we resolved RECX-1 into a binary
with a separation of 0\farcs141 and a position angle of 21$^\circ$
(taking into account the fact that MIRRORB creates spurious images),
indicating some relative motion between the April 2004 $H$-band images
of \citet{brandeker06} and our January 2010 observations.
\citet{brandeker06} found that the semi-major axis of the orbit was
0\farcs42.  The two stars are also marginally resolved in our STIS
spectra, which were obtained with a slit position angle of
-49.6$^\circ$.  Separate spectral extractions for the components were
obtained by fitting two Gaussians to the cross-dispersed profile at
each wavelength position, subtracting off the emission from one
component, and subsequently extracting the counts in the other
component.  From the blue spectrum, we estimate spectral types of K4
for the S component and K6 for the N component, consistent with the
unresolved spectral type of K5--K6 \citep{luhman04}.  The near-UV
spectra of the two stars are similar, except for a $\sim$ 1.5 times
stronger
\ion{Mg}{2} $\lambda2800$ flux from the S component.  For our
analysis, we use the combined spectra of the N and S components to
increase the signal to noise.  With both components having spectral
types close to RECX-11 and very similar STIS spectra, little error is
introduced by using the combined spectra.

\subsection{Ground Based Observations}
Both sources were observed using MIKE (Magellan Inamori Kyocera
Echelle) on the Magellan-Clay telescope at Las Campanas Observatory in
Chile \citep{bernstein03} on 11 March 2010, with a coverage of
4800--9000 {\AA} and resolution, R$\sim$35,000.  The data were reduced
using the Image Reduction and Analysis Facility (IRAF) tasks CCDPROC,
APFLATTEN, and DOECSLIT \citep{tody93}.

We also obtained 4 low dispersion spectra of RECX-11, covering the
H$\alpha$ line, between 27 November and 22 December 2009, using the
Small and Medium Aperture Research Telescope System SMARTS 1.5m
telescope at CTIO. We used the RC spectrograph, which is a long-slit
spectrograph with a Loral CCD detector; the slit subtends 5~arcmin on
the sky and has a 1.5~arcsec slit width.  All observations were made
by SMARTS service observers and each night consisted of 3 observations
with 5 minute integration time.  The three observations are
median-filtered to minimize contamination by cosmic rays. The data was
reduced with the SMARTS spectroscopic data reduction
pipeline\footnote{http://www.astro.sunysb.edu/fwalter/SMARTS/smarts\_15ms
  ched.html\#RCpipeline}. We subtracted the bias and trimmed the
overscan and flattened the image using dome flats. The spectra were
extracted by fitting a Gaussian plus a linear background at each
column.  Uncertainties were based on counting statistics, including
uncertainties in the fit background level.  The wavelength calibration
was based on an arc lamp spectrum obtained before each set of images.

Finally, we obtained $R=100,000$ spectra of the \ion{He}{1}
$\lambda10830$ line from RECX-11 with Phoenix on Gemini South on 12
December 2009 and from both RECX-1 and RECX-11 using CRIRES on VLT-UT1
on 22 May 2011.  The observations were obtained in an ABBA nod
pattern.  The star $\eta$ Cha, a B8 pre-main sequence star with no
detectable photospheric features at 10830 \AA, was observed to provide
the telluric correction. These spectra were reduced with
custom-written routines in IDL. The wavelength solution is accurate to
$\sim1$ km s$^{-1}$ and was calculated from a linear fit to telluric
absorption lines obtained from spectra generated by ATRAN (Lord 1992)
and available online.

\section{Accretion Analysis}
\label{evidence}
Figure \ref{stis} shows the 
STIS NUV spectra of RECX-1 and RECX-11; they are remarkably 
similar, making it difficult to assess if either is accreting based
on a UV excess.  Also shown in Figure \ref{stis} is a STIS spectrum of the
inactive K5V star HD 154363 
\citep[][properties in Table \ref{tabprop}]{martinez10} 
taken from the HST/STIS Next
Generation Spectral Library \citep{heap07}, for comparison.  Both
$\eta$ Cha sources have excess NUV emission compared to the main
sequence standard (\S \ref{chromosphere}).  
Despite the similarity of NUV fluxes,
analysis of emission line profiles 
shows that one source, RECX-11,  is still accreting material from the disk
while the other, RECX-1,  is not. 
In this section we review this evidence. We find that
the emission line profiles of RECX-11 show redshifted
absorption components indicative of accreting material, and
consistent with magnetospheric accretion. 
As mentioned before, accretion shock emission is then expected
from the accreting star, but it must be hidden by 
intrinsic chromospheric emission present in both the accreting and
the non-accreting stars; this points to 
a very low accretion luminosity and mass accretion rate. 
In \S 4
we estimate upper limits for these quantities.

\subsection{Diagnostics of Low Accretion}  

\subsubsection{Lines Produced by Accretion and Related Processes}

\begin{figure}
\plotone{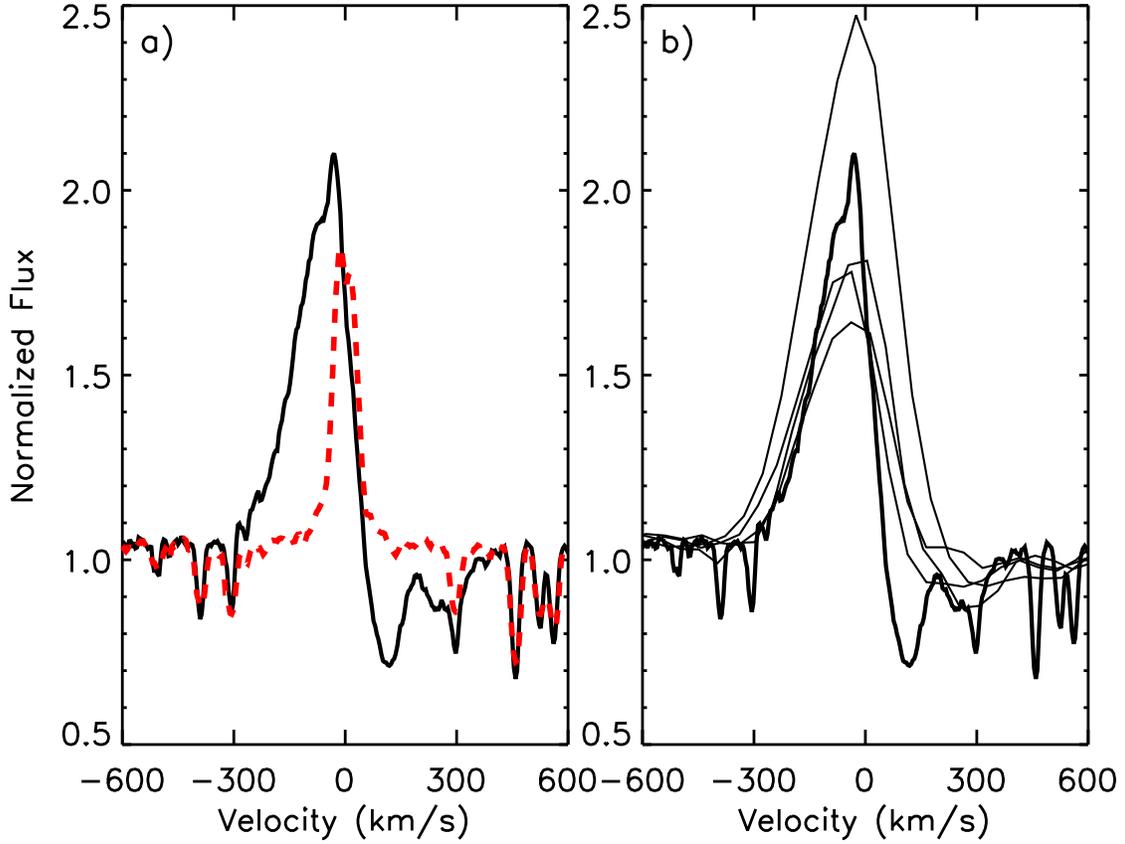}
\caption{H$\alpha$ line profiles. (a) MIKE spectra of RECX-11 (solid line)
and RECX-1 (red dashed line). (b) SMARTS spectra of RECX-11 obtained within a few days of the
HST observations. The MIKE spectrum is also shown for comparison as the thick solid line.
The wide blue wing and red-shifted absorption in the H$\alpha$
profile of RECX-11, characteristic of accretion, are observed
in the MIKE and SMARTS data. Transient narrow red-shifted components
are observed in the red wing of the lines and show the largest
variability. 
}
\label{halph}
\end{figure}

Our spectroscopic observations, from the FUV to the IR, include
a number of emission lines that trace phenomena such as accretion
and outflows in T Tauri stars. 
Line widths of several hundred $\rm {km \, s^{-1}}$ are used
to identify accretors, but 
the most conspicuous 
indicators are redshifted and blueshifted
absorption components on the emission line profiles.
  
Accessible in the optical, the H$\alpha$ emission
line is one of the most commonly used accretion tracers.  
Figure \ref{halph} shows 
MIKE H$\alpha$ line profiles for RECX-11 and RECX-1.
The H$\alpha$ equivalent widths (EW) of RECX-11
and RECX-1 are 4 {\AA} and 1.3 {\AA}, respectively, which would make
RECX-1 a WTTS and RECX-11 a borderline CTTS/WTTS according to the
standard criterion for a K5--K6 star \citep{white03}.  The H$\alpha$ profile of
RECX-1 in Figure \ref{halph}a agrees with the WTTS classification
because it is symmetric and narrow.  With a line width at 10\% of the
peak intensity of only $\sim$130 km s$^{-1}$, the line shows no
indication of the high velocities that are characteristic of
magnetospheric accretion.  In contrast, the H$\alpha$ profile of
RECX-11 is typical of accreting sources, with a line width at 10\% of
$\sim$300 km s$^{-1}$ \citep{white03}.  The blue emission wing is
wide, extending to velocities of several hundred km s$^{-1}$, implying
that material is accreting at nearly free-fall velocities.  Inverse P
Cygni absorption due to infalling material along the
accretion streams is also seen \citep{lima10,muzerolle03,walter99}.
The line profile is consistent with model predictions for a high
inclination source \citep{muzerolle01,kurosawa06}, in agreement with
the inclination of {\it{i}} $\sim68^{\circ}$, estimated from 
the values of the projected rotation velocity
$v$sin$i$ \citep{jay06} and the rotational period \citep{lawson01}.
Moreover,
\citet{lawson04} compared their observed H$\alpha$ profile of RECX-11 with 
magnetospheric accretion models and obtained a good fit
assuming an inclination of {\it{i}} $\sim70^{\circ}$.

In Figure \ref{halph}b, we show our MIKE profile and the four SMARTS
low resolution H$\alpha$ profiles of RECX-11 obtained within a few
days before and after the HST observations.  The H$\alpha$ profile is
highly variable, as already noted by \citet{lawson04} and
\citet{jay06}.  The wide blue emission wing is present in all spectra,
but variability is conspicuous in the red wing, sometimes showing
narrow transient red-shifted absorption components.  This variability
points to a complex geometry of the accretion streams very close to
the star since red-shifted absorption occurs when material in the
accretion columns absorbs hot radiation produced in the accretion shock
on the stellar surface
\citep{bouvier07,muzerolle01}.  
The complex structure revealed by
the line profiles of H$\alpha$, 
which in the case of
RECX-11 is conclusive in identifying accretion,
would have been missed 
when only considering the equivalent width of the line.

\begin{figure}
\plotone{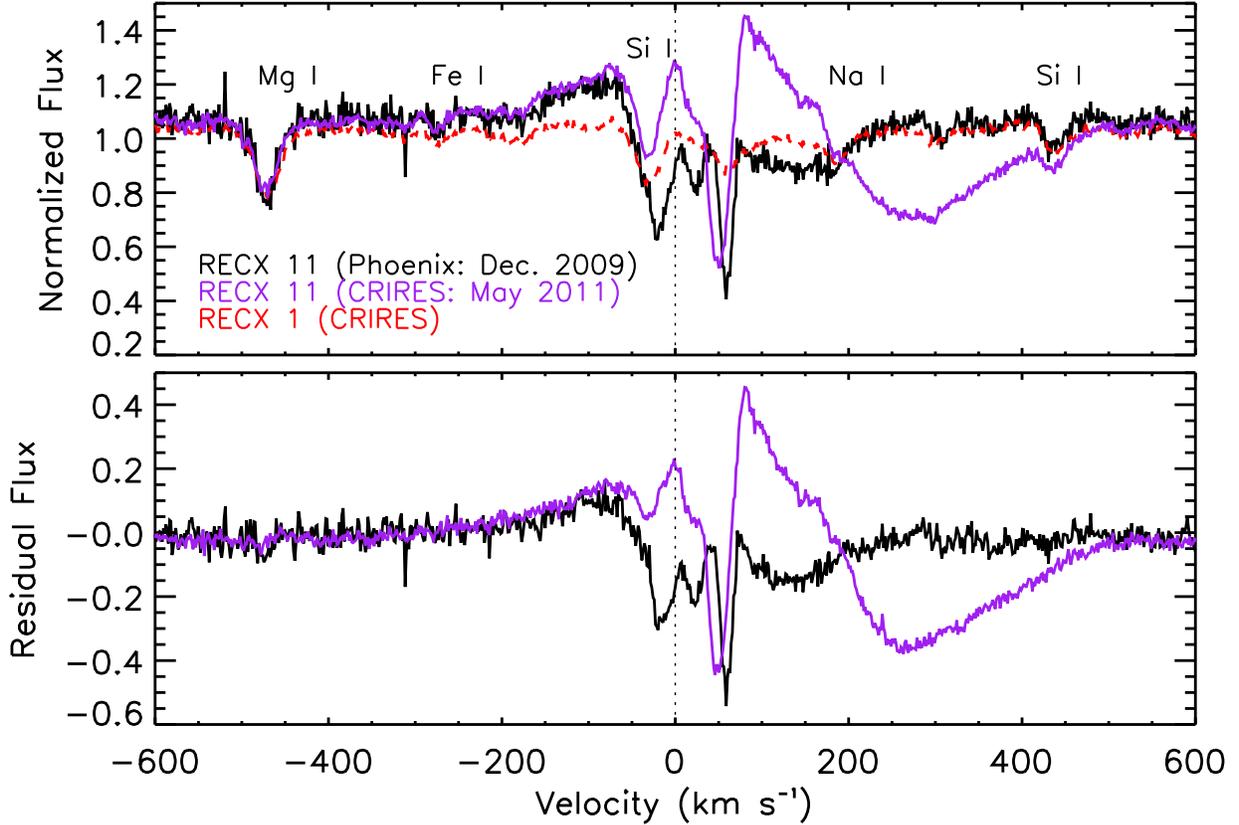}
\caption{He I $\lambda$10830 profiles of RECX-1 and RECX-11.
{\emph{Top Panel:}} He I observations of RECX-11 are shown from two epochs largely separated in time by 18 months.  RECX-1 is shown for comparison and photospheric lines identified by the Infrared Telescope Facility (IRTF) spectral library are designated \citep{rayner09}.  RECX-11 shows significant variability between the two observations, including a variable redshifted emission component and a broad, high velocity redshifted absorption component characteristic of magnetospheric infall.
{\emph{Bottom Panel:}} He I $\lambda$10830 line profiles of RECX-11 after subtraction of the RECX-1 line profile.  The Phoenix observation was degraded to the resolution of the CRIRES observations prior to subtraction of the RECX-1 profile.  The photospheric lines are cleanly subtracted but the complex nature of the profile remains.
}
\label{heI}
\end{figure}

In Figure \ref{heI} we show two He I
$\lambda$10830 profiles for RECX-11 obtained 18 months apart
with Phoenix and CRIRES along with a CRIRES spectrum of RECX-1 for
comparison.  We also show the RECX-11 line profiles
after subtraction of the RECX-1 line profile (with the Phoenix
observation at the resolution of CRIRES) which removes any
contribution by photospheric absorption lines.  
The He I
$\lambda$10830 emission profiles show absorption components
similar to those found by
\citet{edwards06} and \citet{fischer08} 
in a large sample of CTTS,  
which were interpreted as arising in the wind
and in the accretion columns.  
The earlier Phoenix
profile shows a broad red-shifted absorption component extending to
$\sim 250\; {\rm km \, s^{-1}}$ with several deep, narrow red and
blue-shifted absorption lines at lower velocities, between $-50\;
\rm{and}\; 100 \; {\rm km \, s^{-1}}$.  The second, CRIRES, profile
also shows a complex absorption spectrum with a broad component
extending from 200--400 km s$^{-1}$.  It also shows a red-shifted
emission wing which was not observed in the Phoenix observation.
The broad high velocity absorption feature in the
CRIRES spectrum occurs at velocities typical of material infalling
along accretion streams.  According to magnetospheric accretion
models, material falling from 5$R_{\ast}$ in the disk onto a star with
the mass and radius of RECX-11 will reach velocities of
$\sim470\;\rm{km\;s^{-1}}$, capable of producing the absorption
features at the observed velocities \citep{calvet98}.  

\citet{fischer08} found that sources with low accretion rates,
observed to have low veiling at 1 $\mu$m ($R_Y<0.5$), were more likely
to have redshifted absorption in He I than sources with high accretion
rates.  \citet{fischer08} also found that in order to explain the He I
$\lambda$10830 profiles, absorption over a large range of velocities
was necessary.  They proposed that in
low $\mdot$ sources 
accretion occurs in low density, narrow accretion ``streamlets'' with
small filling factors, covering a large velocity range,
although none of their sources had low enough accretion
rates to test this scenario.  The narrow
red-shifted absorption features in
the Phoenix spectrum 
of our very low
accretor, RECX-11 
(\S \ref{smallmdot}), 
may be
observational evidence for these multiple, narrow accretion streams.
Some of these narrow features, 
like the one at $\sim 100
\kms$, may be related to the feature at comparable velocity in
seen in the 
MIKE H$\alpha$ profile
(Figure \ref{halph}), although no definite conclusion can be
stated since the lines were observed at different times.
Simultaneous extensive
coverage of H$\alpha$ and He I $\lambda$10830 is necessary to
understand the complexity of the accretion flows in RECX-11.

\begin{figure}
\plotone{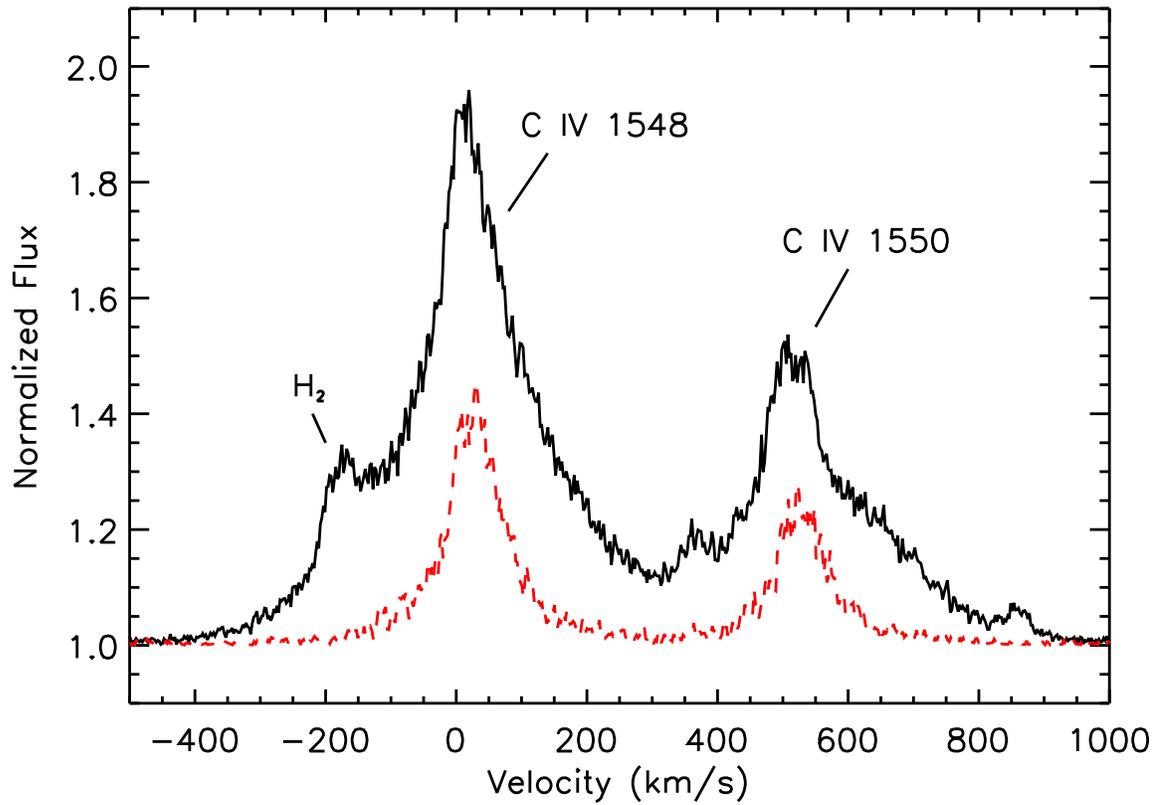}
\caption{
COS C IV $\lambda$1549 and
$\lambda$1551 line profiles for RECX-11 (solid line) and
RECX-1 (red dashed line). In agreement with other accretion 
indicators,  the CTTS RECX-11 has broader lines than the WTTS RECX-1;
the emission is expected to arise in the accretion shock and flows.
}
\label{civ}
\end{figure}

In addition to H$\alpha$ and He I $\lambda$10830, the C IV doublet in
the FUV spectrum can be used to separate accretors from non-accretors.
The C IV $\lambda$1549 line emission has been shown to correlate with
the mass accretion rate, indicating that its origin is, at least in
part, in the accretion shock and flows \citep{johnskrull00,calvet04}.
In Figure \ref{civ} we compare the C IV line profiles of the accretor,
RECX-11, to the non-accretor, RECX-1.  The C IV line profile is much
wider and stronger in the accreting source than in the WTTS, in which
the emission is expected to be chromospheric.  The strength and width
of C IV in RECX-11 is consistent with the accreting nature of RECX-11
(Ardila et al. in preparation).

In the NUV range, the only significant difference between
the spectra of RECX-11 and RECX-1 is  
in the CII] $\lambda$2325 {\AA} line, which is much stronger in RECX-11 (Figure \ref{stis}).
This line has been
observed in other accreting sources \citep{gomez05,calvet04}, and
its likely origin is the accretion shock.

\subsubsection{The \h2 Lines}

The FUV spectrum of RECX-11 contains a wealth of lines due to
molecular species, including \h2 \citep{france11}.  Lines from \h2 in
the UV are mainly due to two mechanisms.  Ly$\alpha$ emission excites
\h2, producing a fluorescent spectrum in the UV as the electrons
cascade back to the ground electronic state
\citep{herczeg02,herczeg04,yang11}. In addition, fast electrons can
collisionally excite \h2, resulting in UV lines and continua
\citep{bergin04, ingleby09}.

Using emission from collisionally excited \h2 measured in low
resolution FUV spectra of a large sample of both accreting and
non-accreting young stars, \citet{ingleby09} showed that by the time
accretion ends, the inner disks are depleted of gas.  Our targets
confirm this trend at high resolution.  In Figure \ref{ctts_wtts} we show the COS FUV spectrum of RECX-11 compared to RECX-1; the wavelengths at which \h2 pumped by Ly$\alpha$ may emit are labeled.  \h2 emission lines at 1434.1 and 1445.3 {\AA} may also be pumped by the C III $\lambda$1174 {\AA} multiplet \citep{herczeg02}.  Figure \ref{ctts_wtts} shows
that the COS FUV spectrum of RECX-11 is rich in lines of molecular
gas, similar to previously observed sources with confirmed accretion
\citep{bergin04,herczeg02,herczeg04,france10}.  
In contrast, 
molecular lines are not apparent in the FUV spectra of the WTTS RECX-1.
If present, the flux of the 
\h2 lines in RECX-1 is $<$5\% of the \h2 flux 
in RECX-11.

Assuming that the molecular lines are formed in the atmosphere of the
disk \citep{herczeg04}, we constructed synthetic models for an
optically thin line formed between radii $R_1$ and $R_2$ in a
Keplerian disk. We assumed that the emissivity depended on radius as
$R^{\gamma}$ and that the intrinsic line profile was a Gaussian with
width corresponding to a temperature of 2500 K.  The intensity at each
radius was convolved with the rotational profile corresponding to the
Keplerian velocity of the annulus, and the flux was then calculated by
integrating the intensity
over the range of radii. Finally, the resultant profile was convolved
with the G130M line spread function.

\begin{figure}
\plotone{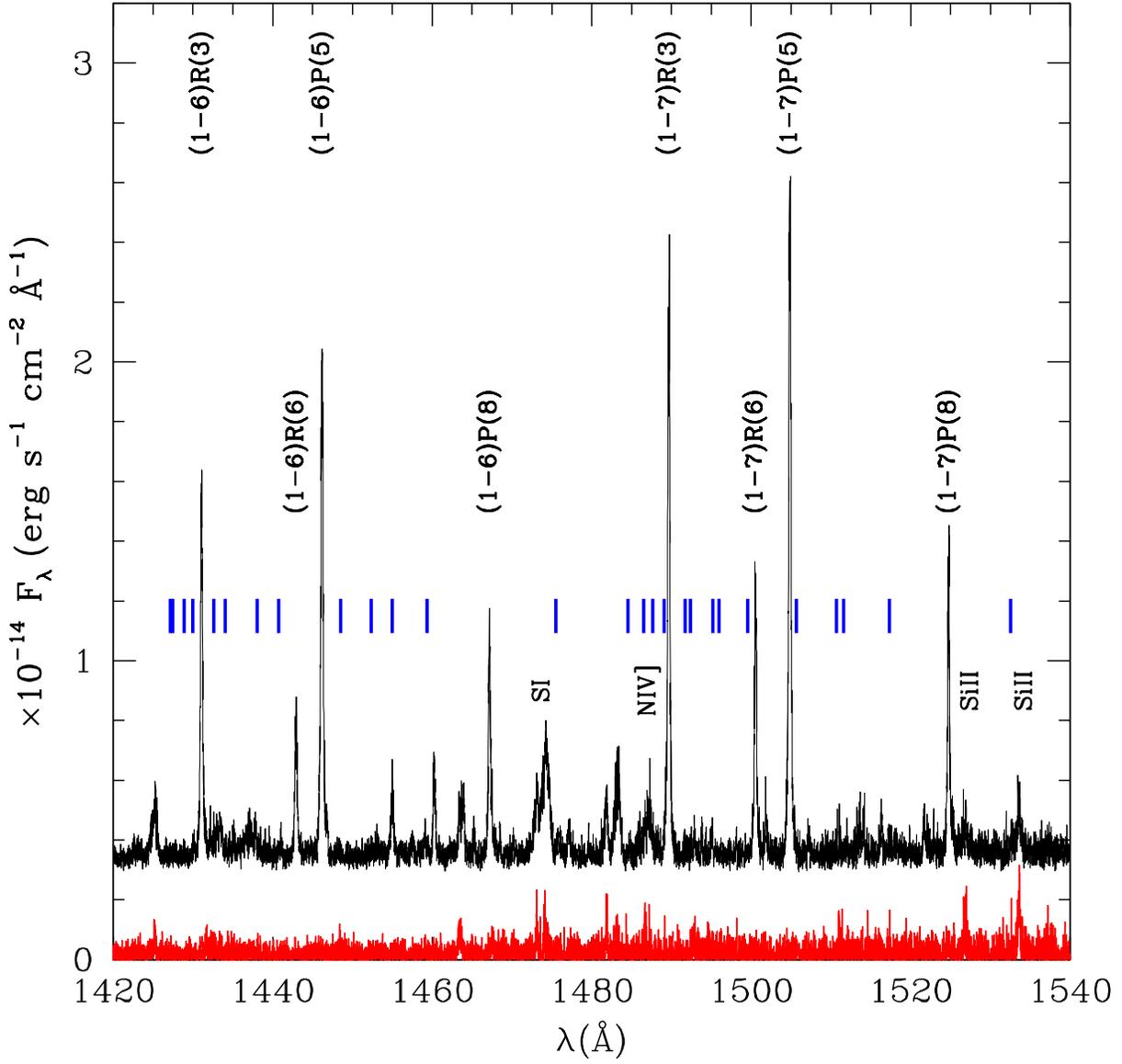}
\caption{COS FUV spectra of the WTTS RECX-1 (lower red) and the CTTS
RECX-11 (upper black).  The RECX-11 spectrum is offset vertically for
clarity.  Strong lines of Ly$\alpha$ fluoresced \h2 and atomic lines
are labeled.  The blue dashes show the wavelengths at which additional
\h2 emission lines from Ly$\alpha$ pumping may be seen.  RECX-1 shows
a lack of molecular gas lines which are apparent in RECX-11;
essentially no gas exists in the inner disk region of RECX-1. }
\label{ctts_wtts}
\end{figure}

\begin{figure}
\plotone{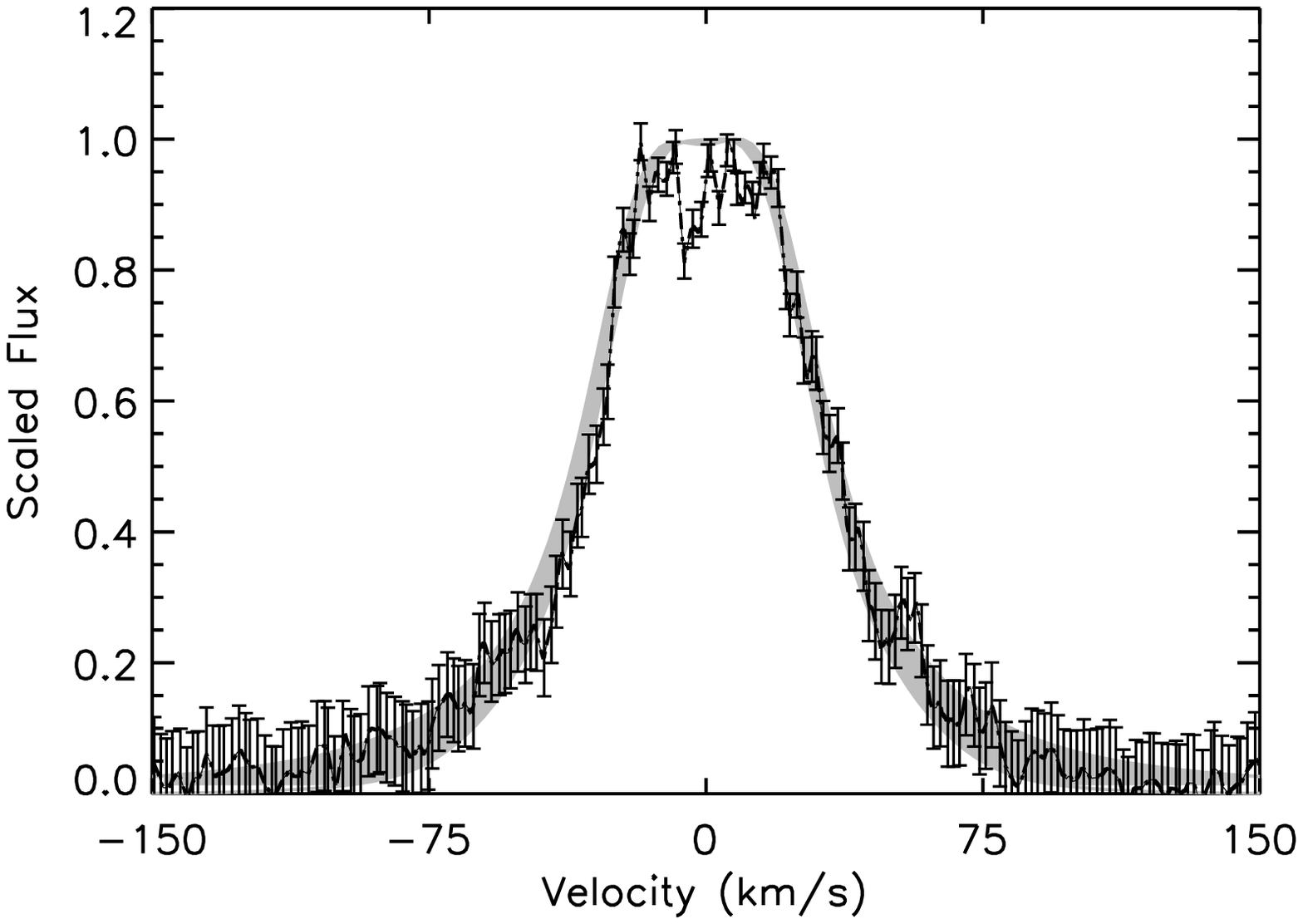}
\caption{Spectrally resolved \h2 line profile.  The dot-dashed line
shows the flux weighted average of four \h2 lines observed with the COS
G130M grating.  These lines result from the (1--2) R(6) and (1--2)
P(5) Ly$\alpha$ pumping transitions and are found in a narrow
wavelength region observed with a single grating. The shaded region
shows the predicted line emission originating in a Keplerian disk with
inclination of 68$^{\circ}$, inner radius $R_1=1-20\;R_{\ast}$ and
outer radius, $R_2=450-500\;R_{\ast}$.  The predicted line fluxes are
convolved with the COS line spread function to account for
instrumental effects.}
\label{h2}
\end{figure}

Figure \ref{h2} compares the predicted profile to a flux-weighted
average of \h2 lines, selected from the spectral region ($<$1330
{\AA}) in which all the lines were observed with the same grating,
G130M.  The continuum of the predicted profile has been scaled to the
observed continuum.  The shaded region in Figure \ref{h2} corresponds
to predicted profiles for models with inner radius between $R_1
=1-20\;R_{\ast}$ (0.01--0.1 AU), outer radius $R_2 = 50-500\;R_{\ast}$
(0.3--3.2 AU), and emissivity power law exponent $\gamma = -1.5$.  $\chi^2_{red}=3.1-3.6$ for the range of models shown in Figure \ref{h2}.  We
find that the inner radius is well constrained, within the
uncertainties of the observations; small inner radii with high
Keplerian velocities are needed to account for the width of the
observed wings.  The outer radius $R_2$, on the other hand, is not
well constrained, since little emission comes from those regions.  The
power law of the emissivity is constrained by the width of the line
near the peak, and we require it to fall rapidly with radius.
This rapid fall-off
is likely due to the combination of a number of factors.  For
instance, the excitation will decrease as the flux of Ly$\alpha$
decreases with distance; similarly, the number of electrons in excited
states that can absorb Ly$\alpha$ transitions decreases with radius,
as the temperature of the upper disk layers drops
\citep{meijerink08}. 
Our modeling indicates that
the region of formation of the \h2 lines can be as close to the star
as 0.01-0.1 AU, as was also found for the accreting brown dwarf 2M1207
\citep{france10}.  This indicates that gas likely extends close to the stellar surface, inside the dust truncation radius, and reaches the stellar magnetosphere, see section 5.2.

\subsection{Unreliable accretion diagnostics at low $\mdot$}
Additional tracers of accretion present in our dataset, 
like [O I] at 6300 and 6363 {\AA} and the Ca II infrared
triplet, do not provide conclusive indication of accretion in
RECX-11.
[O I] emission, which is expected to originate in
an outflow, is observed only in sources undergoing accretion
\citep{hartigan95}.
We show in 
Figure \ref{oi}a  
the line profiles of [O I] $\lambda$6300 of 
RECX-11 and RECX-1.
[O I] $\lambda$6300 appears in RECX-11 in excess
over the WTTS spectrum of RECX-1, but the detection is
$<3\sigma$, and therefore  not reliable.   
Similarly, the Ca II $\lambda$8542 
is often used as an indicator of accretion, and even
used to measure mass accretion rates \citep{muzerolle98}.
However, the profile of this line in
RECX-11,
shown in Figure \ref{oi}b, is indistinguishable from the same line in
RECX-1, indicating that the flux comes mostly from the chromosphere.

The Mg II $\lambda$2800 line luminosity also correlates with accretion
luminosity in accreting stars \citep{calvet04} and is seen in our
STIS spectra (Figure \ref{stis}).  The line is
observed to have the same strength in the weakly accreting star, the
WTTS and the dwarf star, indicating that it is mainly due to
chromospheric activity. 
Moreover, its strength does not increase with
the NUV emission level of the chromosphere.  This agrees with the
results of \citet{cardini07}, who show that at ages $<$0.3 Gyr, the Mg
II $\lambda$2800 fluxes are not observed to change due to saturation effects, 
and so 
the line strength does not reflect the level of chromospheric activity.

\begin{figure}
\plotone{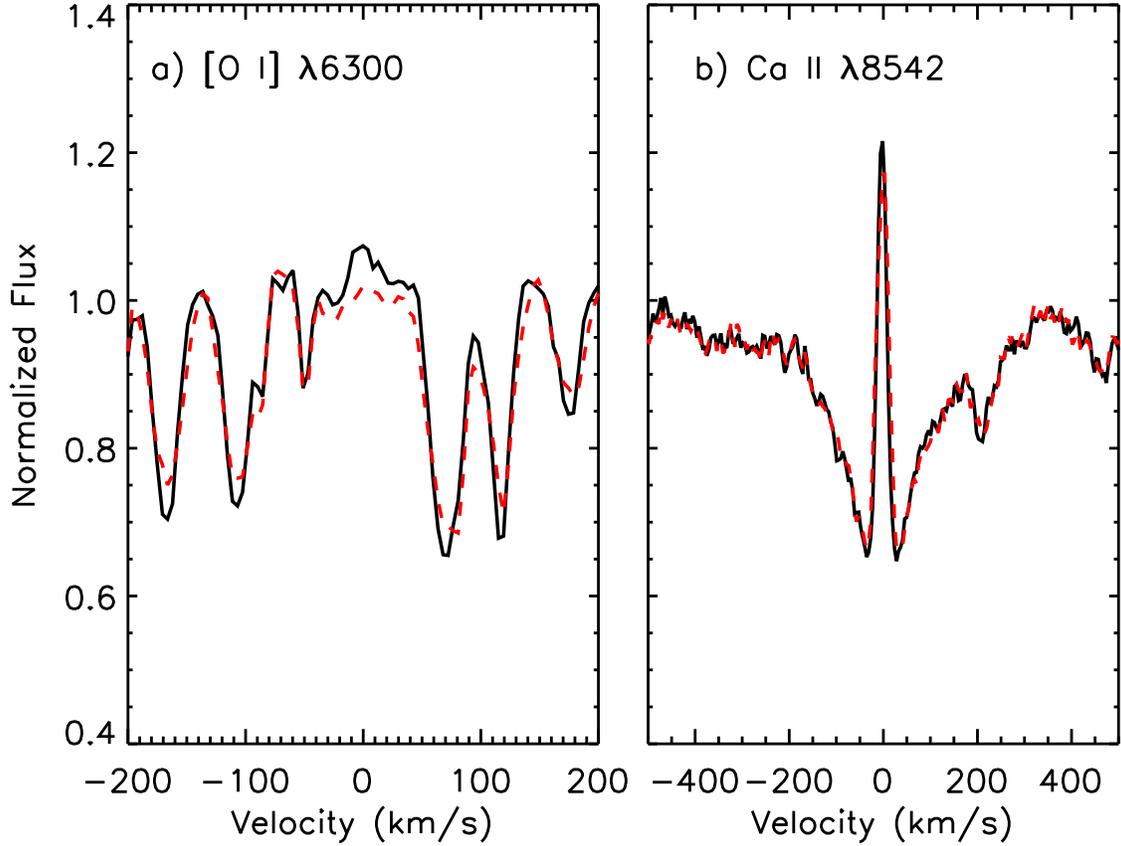}
\caption{Accretion indicators.  In each panel the
black solid line shows the line profile of the CTTS, RECX-11, and the
red dashed line shows the WTTS, RECX-1. a) [O I].  The [O I]
$\lambda$6300 {\AA} line probes wind emission in T Tauri stars.  Weak [O I] emission appears to be present in RECX-11 in excess over the WTTS but the detection is $<3\sigma$.  If [O I] were present it would be additional support for accretion in RECX-11.  b) Ca
II. The Ca II $\lambda$8542 {\AA} shows an emission core on the
photospheric absorption line.  At low $\mdot$, this line
is dominated by chromospheric emission and cannot be used to measure
accretion rates, as shown by the identical profiles of RECX-1 and RECX-11.
}
\label{oi}
\end{figure}

\section{Measuring Small Accretion Rates}
\label{smallmdot}
\subsection{The Chromosphere}
\label{chromosphere}
While dwarf or giant stars have been used to estimate the intrinsic
stellar emission in CTTS when measuring the UV excess, WTTS are
the best templates for accretion analysis because they have similar
surface densities and are expected to have comparable levels
of chromospheric emission. 
The presence of a UV excess in
non-accreting pre-main sequence stars and active stars (when compared
to inactive dwarf stars) due to enhanced chromospheric activity is
well established \citep{houdebine96,franchini98}.  In fact, recent studies
with GALEX have relied on this excess to identify young stars in
nearby star forming regions
\citep{rodriguez11,shkolnik11,findeisen10}.  
This UV excess due to chromospheric emission affects
measurements of the excess emission from which 
accretion luminosities are derived.
Several studies have already taken these effects into account.
For instance,\citet{valenti03} used
the average spectrum of several WTTS 
to estimate the UV excess in low resolution International Ultraviolet
Explorer (IUE) spectra of accreting stars.  Similarly,
\citet{herczeg08} 
used WTTS templates to measure flux excesses and
estimate accretion rates in
very low mass objects.
However, many determinations of accretion luminosities and accretion
rates based on UV fluxes 
have used dwarf standards as templates \citep{gullbring00,calvet04}. The omission of 
the chromospheric fluxes 
makes these determinations, and calibrations of line luminosities
derived from them, uncertain at the low end.

In this work we have presented UV fluxes of RECX-1, which 
to date, is the only WTTS spectrum with the signal
and resolution needed for use as a chromospheric template.
We compare the spectrum of the inactive K5 main sequence star (HD
154363) with that of RECX-1, shown in Figure \ref{stis}.
The flux between 2000--5000 {\AA} from RECX-1 is 0.05 $\lsun$ brighter
than that from HD 154363.  If this luminosity difference were due to
accretion, we would have estimate that RECX-1 is accreting at log $\mdot>
-8.5\;\msunyr$, which is typical of a CTTS.  This is a lower limit because we
have not accounted for any excess emission outside of the range of
observations. This clearly 
demonstrates the effect of an active chromosphere on the UV spectrum
and shows that inactive main sequence standards should not be used to
measure small NUV excesses.  It can be argued that the chromospheric level of RECX-1 is not
typical of WTTS because it is a binary.  However, with a binary
separation $>$ 15 AU \citep{kohler02,brandeker06} it is unlikely
that the stars are interacting.  In any event, a survey of NUV spectra
of WTTS of different spectral types and luminosities is needed to
determine the characteristic range of chromospheric NUV emission in
these stars.

\subsection{Estimate of the mass accretion rate from NUV emission}
\label{shock}
With evidence that RECX-11 is accreting, we use 
shock models to estimate upper limits
to the accretion luminosity and mass accretion rate of RECX-11.
A full description of the accretion shock models can be found in
\citet{calvet98}; here we review the main characteristics.  An
accretion shock forms at the stellar surface when material from the
disk falls onto the star along magnetic field lines (\S\ref{intro}).
The shock formed at the base of the accretion column slows the
material to match the stationary photosphere, converting the kinetic
energy into thermal energy, and causing the temperature to increase
sharply.  The shock emits soft X-rays into the pre-shock
and post-shock regions and the photosphere below the shock. The heated gas
emits the observed excess continuum emission which is strongest in the
UV \citep{calvet98}.
The accretion shock model is used to fit the NUV excess emission and
account for shock emission outside of the range of observations.  
The model yields the accretion luminosity, from which the
mass accretion rate can be derived with knowledge of the stellar
mass and radius, through $L_{acc} = G M \mdot / R$.
The accretion shock model assumes that all the excess flux
in the UV comes from
accretion, neglecting any contribution from the outflow.
Since 
the luminosity of the wind is $<$10\% of the accretion luminosity
for typical mass loss rates
\citep{cranmer09} 
this assumption is justified.

We model the spectrum of RECX-11 as the sum of the WTTS RECX-1, used as
a proxy for the intrinsic stellar emission, and the
accretion shock emission; our fit to the observations is shown in Figure \ref{RECX-11} and has a $\chi^2_{red}\sim11.5$. 
Since the NUV fluxes of RECX-11 are essentially chromospheric,
our aim is to determine
the amount of shock emission that may
be hidden in the spectrum of RECX-11.  We find that $\mdot
\sim 3\times10^{-10}\;\msunyr$ is the highest value of the mass
accretion rate that RECX-11 could have and still not show an excess
over the chromospheric emission or veiling in the optical absorption
lines.

\begin{figure}
\plotone{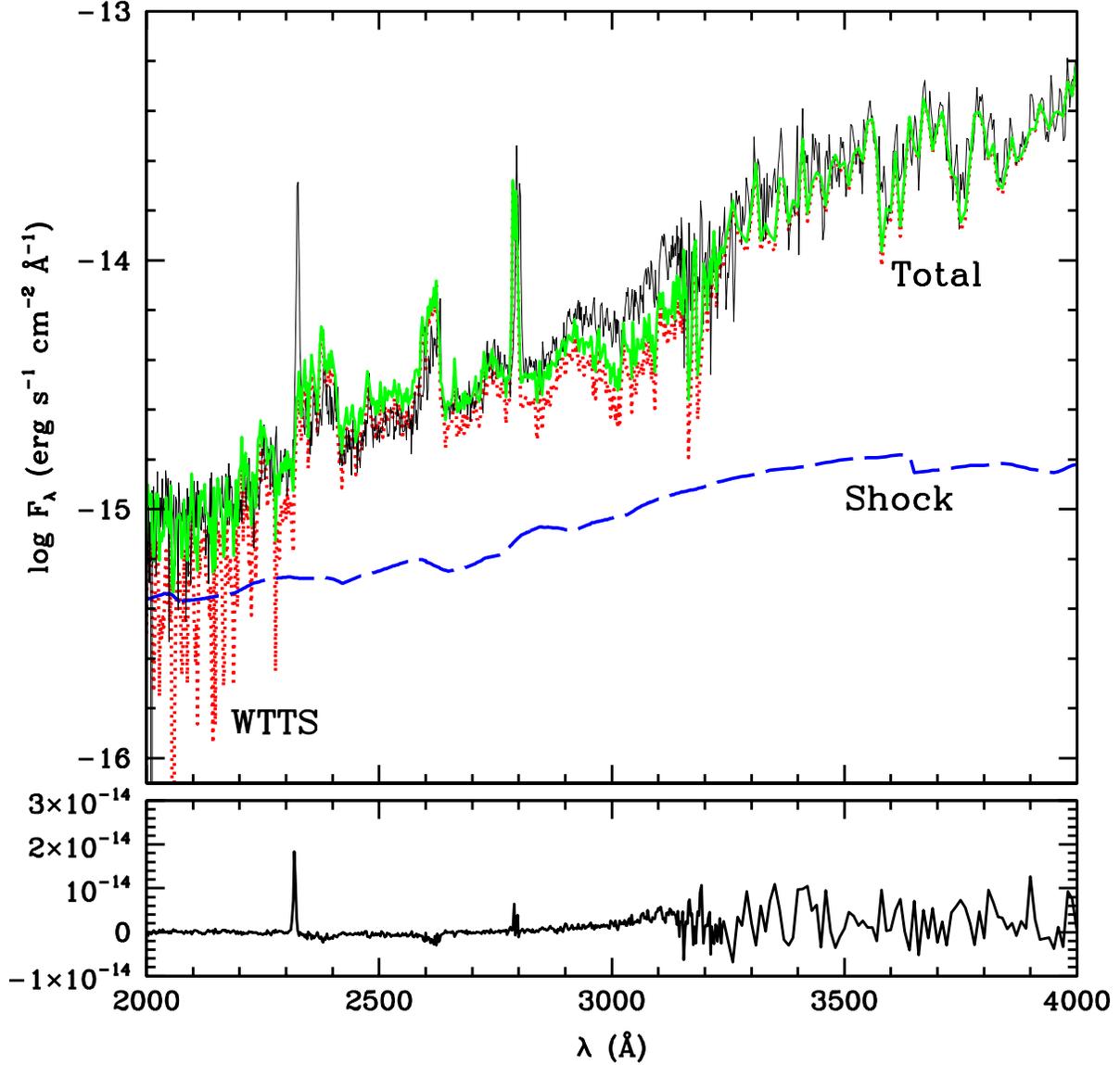}
\caption{STIS spectrum of RECX-11 compared to predictions of the
accretion shock model.  {\it Top Panel:} We show the STIS spectrum of RECX-11 (thin
solid line) along with the template NUV and optical emission, the WTTS
RECX-1 (red dotted line), and the emission predicted from the
accretion shock model (blue long dashed line), with
$\mdot\le3\times10^{-10}\;\msunyr$.  The thick solid green spectrum is
the accretion shock model added to the WTTS spectrum.  {\it Bottom Panel:} The model (green) was subtracted from the observations (black) and the residuals are shown.
}
\label{RECX-11}
\end{figure}

The upper limit on $\mdot$ is consistent with estimates of accretion
from the emission lines shown in Figures \ref{halph} and \ref{civ}.
From the width of the H$\alpha$ line at 10\% of the maximum and the
relation between 10\% width and $\mdot$ of \citet{natta04}, we find
$\mdot=2.3\times10^{-10}\;\msunyr$.  Also, from the flux in the C IV
$\lambda$1549{\AA} line, we find $\mdot=1.6\times10^{-10}\;\msunyr$
from the relation between C IV flux and $\mdot$ in \citet{valenti03}.
\citet{lawson04} fit their H$\alpha$ line profile of RECX-11 with a
magnetospheric accretion model, finding good agreement for a model
with a mass accretion rate of $\mdot = 4 \times 10^{-11} \msunyr$,
also consistent with our upper limit.

\section{Discussion}
\label{discussion}
\subsection{$L_{acc}$ vs $L_U$}
\label{lacclu}

While it is ideal to measure $L_{acc}$ from the peak of the excess
emission, in the UV, it is often not possible to obtain these
observations.  To overcome
this difficulty, \citet{gullbring98} calibrated the excess luminosity
in the $U$ broad band, $L_U$, in terms of the accretion luminosity,
$L_{acc}$, for a sample of stars for which they had medium resolution
spectra from 3200 to 5400 {\AA}. Accretion luminosities were measured
for each star in this sample by extracting the excess flux through
veiling measurements and using a slab model to account for energy
outside the observed bands. The excess luminosity in $U$ was measured
by subtracting the flux in the $U$ band of a main sequence standard
with the same $V$ magnitude from the observed $U$ band fluxes.  This
correlation was later reproduced using the accretion shock model
described in \S\ref{shock} \citep{calvet98}.  The $L_{acc}$ vs $L_U$
relation is commonly utilized in studies of large populations because
it provides a convenient method for finding $L_{acc}$
\citep{rigliaco11a,ingleby09,grosso07,robberto04,white01,rebull00,hartmann98}.
However, as discussed above, the calibration was developed using dwarf
stars as templates against which the $U$ band excess was measured, and
as we have shown, T Tauri stars have active chromospheres which
contribute to the $U$ excess.

The UV excess observed in the WTTS RECX-1 significantly decreases the
estimated excess from the accretion shock emission in the
NUV. Assuming that the chromospheric emission in RECX-1 is typical for
$\sim$ K5 stars, then the intrinsic chromospheric NUV flux would be
0.05 $\lsun$ (\S \ref{chromosphere}.)  If this NUV excess were
interpreted as an ``accretion luminosity", using the
\citet{gullbring98} relation between $L_{acc}$ and $U$ band excess,
the excess luminosity in the U band just due to chromospheric emission
would be 0.01 $\lsun$. This suggests that for the $\sim$ K5 spectral
range, $U$ band excesses and derived accretion luminosities below
these limits should be taken with caution.  An extensive survey of the
intrinsic chromospheric emission in WTTS covering a wide spectral
range is necessary to quantify the expected chromospheric emission in
CTTS.  A re-calibration of the $L_U$ vs. $L_{acc}$ relation using
characteristic chromospheric contributions may extend its range of
application to low values of $L_{acc}$ and $\mdot$.

\subsection{RECX-11 and Disk Evolution}
\label{transition}

RECX-11 is in the $\eta$ Chamaeleontis group, with an estimated age of
5-9 Myr.  This age range is interesting because disk frequency studies
indicate that $\sim$ 80\% of the original disks have already
dissipated by $\sim$ 5 Myr \citep{hernandez08}.  The disk around
RECX-11 is then one of the few survivors and a study of its properties
can give insight into the mechanisms of disk survival and dissipation.

In the age range of RECX-11, viscous disk evolution would predict a
value for the $\mdot$ between $10^{-10}\; \msunyr$ and $
10^{-8}\; \msunyr$, extrapolating for the dispersion observed in other
populations \citep{calvet05}.  The $\mdot$ for RECX-11 is below
this range, so additional disk dispersal mechanisms like
photoevaporation may be at play.
According to one type of photoevaporation model, EUV photons would
ionize the disk surface; a wind is established outside a radius $r_g$
where the gas escape velocity becomes higher than the sound speed.
The predicted mass loss rates for EUV photoevaporation is
$\sim10^{-10}\;\msunyr$ \citep{alexander07}.  After the mass accretion
rate becomes comparable to the mass loss rate, mass does not reach the
inner disk, which drains onto the star in timescales comparable to the
local viscous time, $\sim 10^5$ years \citep{clarke01}.  An inner disk
hole is thus created, and EUV irradiation of its edge is expected to
quickly erode the rest of the disk \citep{alexander06}.  For the mass
of RECX-11 (Table \ref{tabprop}), the value of $r_g$ \citep{clarke01}
would be $\sim$ 7 AU.  With further refinements of the theory, the
value of the critical radius can be decreased to $\sim 0.1-0.2\; r_g$
\citep{font04,adams04}.  This implies that if EUV photoevaporation were
active with mass loss rates greater than the low accretion rate of
RECX-11, a hole of at least 0.7 AU should have developed in the
RECX-11 disk.

A second type of photoevaporation model includes the effects of
stellar X-ray and FUV emission and predicts higher mass loss rates
than the EUV models \citep{ercolano08,gorti09a}.  According to the
predictions of the X-ray driven photoevaporation models, the wind mass
loss rate scales with the X-ray luminosity.  The X-ray luminosity of
RECX-11 has been recently measured by XMM Newton to be $3 \times
10^{30}\; {\rm erg \, s^{-1}}$ \citep{lopez10}, very similar to the
value $1.6 \times 10^{30}\; {\rm erg \, s^{-1}}$ estimated from ROSAT
observations \citep{mamajek00}.  The disk of RECX-11 must have been
subject to this influx of high energy radiation for most of its
lifetime, since the X-ray luminosity stays approximately constant in
the 1 - 10 Myr age range \citep{ingleby11}.  With the observed value
of the X-ray luminosity, the mass loss rate predicted by X-ray driven
photoevaporation is $\sim 10^{-8} \msunyr$, and the low value of the
accretion rate would imply that the disk was quickly clearing its
inner regions, with timescales $\le 10^5$ yrs \citep{owen11}.  So, for
the X-ray photoevaporation models, the innermost disk regions should
be cleared or in the process of being cleared.

To find evidence of the effects of photoevaporation on the disk of
RECX-11, we have constructed the SED shown in Figure \ref{irs}.  We
plot fluxes at the $B$, $V$, $R$, $I_C$ bands \citep{lyo04,lawson01}
and $J$, $H$, and $K$ bands from the 2MASS survey \citep{skrutskie06}
along with \emph{Spitzer}/IRAC [3.6], [4.5], [5.8], and [8] and MIPS
[24] and [70] bands from \citet{sicilia09}.  We also include the
\emph{Spitzer} IRS spectrum, observed in the SL1 and LL modes; the
spectrum was downloaded from the Spitzer archive and reduced with
version S18.7 of the Spitzer Science Center pipeline using the SMART
data reduction package \citep{higdon04} following the description in
\citet{mcclure10}.  The SED of a K5 star, scaled at $J$, is shown for
comparison; this SED was constructed with colors taken from
\citet{kenyon95}.  The median SED of Taurus \citep{dalessio99}, also
scaled at $J$, is also shown.

\begin{figure}
\plotone{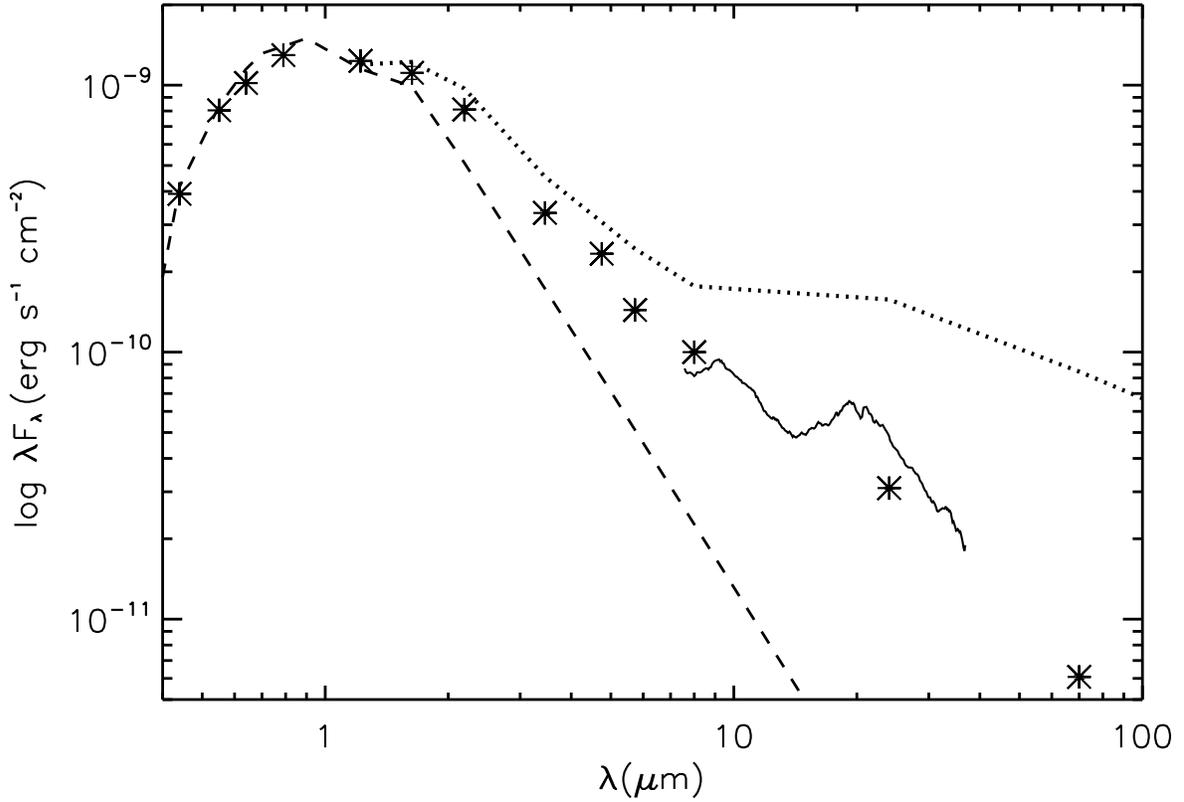}
\caption{IR spectrum of RECX-11.  The IRS spectrum (solid line) and
\emph{Spitzer} photometry (asterisks) show that the IR fluxes are
lower than the median IR fluxes of Taurus (dotted line), yet still in
excess over the photosphere (dashed line) even in the near-IR.  A
near-IR excess is due to dust in the inner circumstellar disk,
extending all the way to the dust sublimation radius. 
}
\label{irs}
\end{figure}

The SED shown in Figure \ref{irs} of RECX-11 shows excess above the
photosphere at all wavelengths $>1.5\;\mu$m.  Beyond the near
infrared, the fluxes are lower than the median SED of Taurus, a proxy
for optically thick disks in which dust is present all the way into
the dust sublimation radius.  The slope of the SED between 13 and 31
$\mu$m of $n_{13-31}=-0.8$ is steeper than the median value of -0.4
for Taurus, indicating a significant degree of dust settling in the
RECX-11 disk \citep{furlan06,furlan09,mcclure10}.  In addition, the
profiles of the 10 $\mu$m and 20 $\mu$m silicate features in the IRS
spectrum are wider than those found in the ISM \citep{sicilia09} and
contain substructure consistent with significant dust evolution,
either in the form of grain growth or crystallization
\citep{watson09}.

Despite the high degree of evolution of dust in the disk, the SED of
RECX-11 does not show indications of an inner cleared region.  In
fact, the location of RECX11 in the K-[8] vs. K-[24] diagram
corresponds to that of full disks according to the study of
circumstellar disks at various stages of evolution by
\citet{ercolano11}.  Moreover, the SED of RECX-11 shows a substantial
near-IR excess, comparable to the median SED of Taurus, despite the
low fluxes at longer wavelengths.  The largest contributor to the
near-IR excess in T Tauri star disks is the optically thick ``wall''
at the dust destruction radius, the sharp transition between the dust
and gas disk \citep{dalessio06, dull10}. For the parameters of
RECX-11, the dust destruction radius is of the order of $\sim 0.06$
AU, taking the simple expression for the equilibrium temperature, with
$T = 1400$ K as the dust sublimation temperature, and $T_{eff}$ =
4350K for a K5 star \citep{dull10}.

The evidence then seems to rule out inner clearing in the disk of
RECX-11. First, it is accreting, so mass has to reach the
magnetospheric radius, 3-5 stellar radii, $\sim$ 0.03 AU. Substantial
\h2 emission is observed, coming from gas regions as close in as
0.01-0.1 AU to the central star, typical of all accretors.  Moreover,
dust extends inwards to $\sim$ 0.06 AU. These inner radii, where disk
material is present, are not consistent with those predicted for
photoevaporation models, according to which the innermost $< 0.1$ AU
of the disk should be clear.  It could be argued that the disk of
RECX-11 has been caught in the short phase when the inner disk is
still accreting its remaining mass onto the star. However, the recent
analysis of the disk of another star in the $\eta$ Cha group, ET Cha
(ECHA J0843.3-7915 = RECX-15) casts some doubt on this suggestion
\citep{woitke11}. The disk of ET Cha is also accreting
\citep{lawson04}; it is very evolved with very low dust and gas
masses, but still has a very substantial near-IR excess consistent
with emission from the dust destruction radius \citep{woitke11}.  For
a 5-9 Myr population of less than 20 stars \citep{mamajek00} the
probability of finding 2 stars in a phase lasting only $\sim 10^5$
years is low.  Overall, the evidence provided by low accretors in
evolved populations suggests that current models of disk dissipation
by photoevaporation with high mass loss rates may need to be
re-visited.  Our upper limit on $\mdot$ does not, however, rule out
photoevaporation at the low mass loss rates predicted by the EUV
photoevaporation models and indeed, evidence for EUV driven mass loss has been observed in several young stars \citep{pascucci09b}.

\section{Summary and Conclusions}
We used multi-wavelength observations of the CTTS RECX-11 and the WTTS
RECX-1 to investigate their accretion properties. We found:

\begin{itemize}
\item
Multi-wavelength observations provide the clearest picture of the
accretion properties of a low $\mdot$ T Tauri star.  Observations of
molecular gas in the FUV may be the most sensitive accretion probe,
since these lines are only present in accreting sources
\citep{ingleby09}, and their line profiles are consistent with an
origin in the disk, very close to the star.  In the optical
and near infrared, the most
sensitive indications of accretion come from high resolution
observations of the H$\alpha$ 
and He I $\lambda$10830 emission lines.
These show
variable and multiple
redshifted
absorption components
which reveal the complex structure of the
magnetosphere in a low $\mdot$ source.
For very low values of $\mdot$, lines normally 
used as accretion indicators such as
the Ca II IR triplet lines, arise mostly in the stellar chromosphere and are
not sensitive to accretion.

\item
The contribution from the stellar chromosphere to the NUV flux can be
substantial and may result in estimates of the accretion luminosity
from $U$ excesses that are significantly higher than the actual value
of $L_{acc}$.  For RECX-1, a K5 star that is not accreting, as
determined from its narrow H$\alpha$ line
profile and from the lack of \h2 lines in the
FUV, its chromospheric NUV excess over the photosphere of a standard
dwarf star would result in incorrect estimates of $L_{acc}\sim0.05\;\lsun$
and $\mdot \sim3\times 10^{-9}\;\msunyr$. For the middle K spectral
range,
estimates of accretion less than these limits based on methods
calibrated with standard dwarf stars should be taken with caution.

\item
When accounting for the active chromospheres of T Tauri stars by using
a WTTS as a template against which to measure 
the NUV excess, we determine a
very low $\mdot$ for RECX-11, $\le3\times10^{-10}\;\msunyr$.

\item
Given the low mass accretion rate, the high X-ray luminosity, and the
advanced age of RECX-11, photoevaporation is expected to be driving
the final evolution of its disk; if this was the case,
the innermost disk
regions, $\le 0.1$ AU should be clear.  However, we find evidence for
gas and dust in the innermost disk of RECX-11, as indicated by ongoing
accretion, the detection of \h2 lines formed in the innermost disk,
and emission from optically thick dust at the dust destruction
radius at levels consistent with other CTTS. This, together with 
similar evidence provided by other slowly
accreting, old disks, suggests that models for photoevaporation-driven
disk dissipation resulting in high values of the
mass loss rate should be revisited.

\end{itemize}

\section{Acknowledgments}
We thank Will Fischer for help reducing the Phoenix data to obtain the
He I line profile.  We thank the SMARTS service for obtaining the
SMARTS spectra.  Stony Brook University is a member of the SMARTS
partnership.  This work was supported by NASA grants for Guest
Observer program 11616 to the University of Michigan, Caltech, Stony
Brook University and the University of Colorado.  Based on
observations made with the NASA/ESA {\it Hubble Space Telescope},
obtained from the Space Telescope Science Institute data
archive. STScI is operated by the Association of Universities for
Research in Astronomy, Inc. under NASA contract NAS 5-26555.  C.E. was
supported by the National Science Foundation under Award No. 0901947.
RA acknowledges support from the Science \& Technology Facilities
Council (STFC) through an Advanced Fellowship (ST/G00711X/1).


\begin{deluxetable}{lcccccccc}
\tablewidth{0pt}
\tablecaption{Properties of RECX-1, RECX-11 and HD 154363
\label{tabprop}}
\tablehead{
\colhead{Object} &\colhead{Class}& \colhead{Spectral Type}& \colhead{Luminosity}& \colhead{Mass}& \colhead{Radius} & \colhead{Distance}&\colhead{Age} \\
\colhead{}&\colhead{}&\colhead{}&\colhead{$\lsun$} &\colhead{$\msun$} & \colhead{$\rsun$} &\colhead{pc}& \colhead{Myr}}
\startdata
RECX-1&WTTS&K5--K6&1.0&0.9&1.8&97&5--9\\
RECX-11&CTTS&K5--K6&0.6&1.0&1.4&97&5--9\\
HD 154363&Dwarf&K5&0.2&0.6&0.8&11&$<$1400\\
\enddata
\tablecomments{Masses for the $\eta$ Cha sources were calculated from the position in the HR diagram, assuming the evolutionary tracks of \citet{siess00} with solar metallicity.  Stellar parameters for HD154363 are from \citet{holmberg09} and \citet{takeda07}.}
\end{deluxetable}

\begin{deluxetable}{lcccc}
\tablewidth{0pt}
\tablecaption{Log of Observations
\label{tabobs}}
\tablehead{
\colhead{Object} &\colhead{RA}& \colhead{DEC}&\colhead{Telescope/ Instrument}& \colhead{Date of Obs} \\
\colhead{}&\colhead{(J2000)}&\colhead{(J2000)}}
\startdata
RECX-1&08 36 56.12&-78 56 45.3&HST/ COS G130M/G160M&2010-01-22\\
&&&HST/ STIS G230L/G430L&2010-01-22\\
&&&Magellan/ MIKE&2010-03-10\\
&&&VLT-UT1/ CRIRES&2011-05-22\\

RECX-11&08 47 01.28&-78 59 34.1&HST/ COS G130M/G160M&2009-12-12\\
&&&HST/ STIS G230L/G430L&2009-12-12\\
&&&Magellan/ MIKE&2010-03-10\\
&&&CTIO/ SMARTS RC Spectrograph&2009-11-26\\
&&&CTIO/ SMARTS RC Spectrograph&2009-11-27\\
&&&CTIO/ SMARTS RC Spectrograph&2009-12-15\\
&&&CTIO/ SMARTS RC Spectrograph&2009-12-19\\
&&&Gemini South/ Phoenix&2009-12-12\\
&&&VLT-UT1/ CRIRES&2011-05-22\\
\enddata
\end{deluxetable}

\end{document}